\documentclass[10pt,aps,prd,reprint,nofootinbib,amsfonts,amssymb,amsmath,preprintnumbers,notitlepage,superscriptaddress,longbibliography,floatfix]{revtex4-1}
\pdfoutput=1
\usepackage[utf8]{inputenc}
\usepackage[british]{babel}
\usepackage[Symbolsmallscale]{upgreek}
\usepackage[protrusion=true,tracking=true,kerning=true,spacing=true,final,babel=true]{microtype}
\usepackage{physics}
\usepackage{xcolor}
\usepackage{graphicx}
\usepackage[final]{hyperref}
\usepackage{nth}

\begin{document}
\title{Estimating the angular power spectrum of the gravitational-wave\\background in the presence of shot noise}

\author{Alexander~C.~Jenkins}
\affiliation{Theoretical Particle Physics and Cosmology Group, Physics Department, King's College London, University of London, Strand, London WC2R 2LS, United Kingdom}

\author{Joseph~D.~Romano}
\affiliation{Texas Tech University, Physics \& Astronomy Department, Box 41051, Lubbock, Texas 79409-1051, USA}

\author{Mairi~Sakellariadou}
\affiliation{Theoretical Particle Physics and Cosmology Group, Physics Department, King's College London, University of London, Strand, London WC2R 2LS, United Kingdom}

\date{\today}
\preprint{KCL-PH-TH/2019-59}

\begin{abstract}
    There has been much recent interest in studying anisotropies in the astrophysical gravitational-wave (GW) background, as these could provide us with interesting new information about galaxy clustering and large-scale structure.
    However, this information is obscured by shot noise, caused by the finite number of GW sources that contribute to the background at any given time.
    We develop a new method for estimating the angular spectrum of anisotropies, based on the principle of combining statistically-independent data segments.
    We show that this gives an unbiased estimate of the \emph{true}, astrophysical spectrum, removing the offset due to shot noise power, and that in the limit of many data segments, it is the most efficient (i.e., lowest-variance) estimator possible.
\end{abstract}

\maketitle

\section{Introduction}
\label{sec:intro}

The first two observing runs of Advanced LIGO~\cite{TheLIGOScientific:2014jea} and Advanced Virgo~\cite{TheVirgo:2014hva} have firmly established gravitational-wave (GW) astronomy as a new means of exploring the Universe.
With an ever-growing number of compact binary coalescences (CBCs) being catalogued~\cite{LIGOScientific:2018mvr,gracedb}, attention is now increasingly turning toward other, as-yet-undetected,
GW observables, such as the stochastic GW background (SGWB).
The SGWB is a persistent, pseudorandom GW signal, formed from the incoherent superposition of many GW sources throughout cosmic history~\cite{Allen:1996vm,Christensen:2018iqi}, which can be searched for by cross-correlating data between multiple GW detectors~\cite{Romano:2016dpx,Smith:2017vfk}.

Unlike analogous backgrounds of electromagnetic radiation [e.g., the cosmic microwave background (CMB)], the SGWB is not attenuated by intervening matter, and may give us observational access to a wealth of exotic early-Universe sources, such as cosmic strings, phase transitions, and inflationary tensor perturbations~\cite{Caprini:2018mtu}.
However, one expects the dominant contribution to the SGWB to be the combined emission from CBCs at much lower redshift.
This \emph{astrophysical} gravitational-wave background (AGWB)~\cite{Regimbau:2011rp,TheLIGOScientific:2016wyq,Abbott:2017xzg} is one of the main observational targets of LIGO and Virgo as they approach design sensitivity.
Since the components of CBCs are the results of stellar evolution, they should reside in galaxies,\footnote{%
    Here we are neglecting the possibility that the black holes may be primordial; in this case, there would also be a SGWB from primordial black hole binaries~\cite{Mandic:2016lcn,Clesse:2016ajp,Wang:2016ana}.}
    and the AGWB should trace the distribution of galaxies throughout the local Universe.
There has therefore been significant recent interest in AGWB
anisotropies~\cite{Contaldi:2016koz,
Cusin:2017fwz,
Jenkins:2018lvb,
Cusin:2018rsq,
Jenkins:2018uac,
Jenkins:2018kxc,
Jenkins:2019uzp,
Namikawa:2015prh,
Namikawa:2016edr}
and associated observational searches~\cite{TheLIGOScientific:2016xzw,LIGOScientific:2019gaw,Renzini:2018nee} and data-analysis methods~\cite{Thrane:2009fp,Gair:2014rwa,Romano:2015uma,Ain:2018zvo,Renzini:2018vkx,Conneely:2018wis,Hotinli:2019tpc}, as these might provide an entirely new probe of galaxy clustering and large-scale structure (LSS).

The AGWB has two immediate advantages over traditional LSS studies using galaxy surveys: first, it automatically has full sky coverage, thanks to the global beam pattern of the GW interferometers; second, it has (in principle, at least) unlimited depth, since galaxies are not selected down to some brightness threshold, but instead contribute to the AGWB at any redshift.\footnote{%
    Note, however, that the nearest galaxies dominate the signal, as the GW flux decays as the inverse distance squared.}

There exists, however, a serious obstacle to using AGWB measurements in this way.
Due to the finite number of CBCs throughout the Universe, and the very short time they each spend in the LIGO/Virgo frequency band, the angular power spectrum of the AGWB is dominated by \emph{shot noise}.
As was first shown in \cite{Jenkins:2019uzp}, this shot noise is orders of magnitude larger than the ``true'' angular power spectrum, completely obscuring the much subtler correlations caused by LSS.

In this article, we present a simple but powerful method for estimating the \emph{true} AGWB angular power spectrum in the presence of shot noise, exploiting the statistical independence of different shot noise realisations at different times.
We begin by reviewing the definition and statistical properties of the AGWB angular power spectrum in Sec.~\ref{sec:angular-power-spectrum}.
In Sec.~\ref{sec:shot} we describe how shot noise affects the angular power spectrum, and describe the hierarchical relationship between the ensemble of shot noise realisations and the ensemble of LSS realisations.
In Sec.~\ref{sec:estimator} we construct an unbiased estimator of the \emph{true}, astrophysical power spectrum, and derive a simple expression for its variance (some details of this calculation are given in Appendix~\ref{sec:4th-moment}).
In an appropriate limit, this estimator is the minimum-variance unbiased estimator (MVUE) for the $C_\ell$'s in the presence of shot noise, which we show in Appendix~\ref{sec:min-var}.

Our discussion focuses on the AGWB composed of CBCs, as this will likely be the first stochastic GW signal to be detected.
However, all of our main results hold for \emph{any} stochastic source that is comprised of a finite number of transient events.

\section{The SGWB angular power spectrum}
\label{sec:angular-power-spectrum}

The SGWB is described by the density parameter
    \begin{equation}
        \Omega\qty(f,\vu*r)\equiv\frac{1}{\rho_\mathrm{c}}\frac{\dd[3]{\rho_\mathrm{gw}}}{\dd{\qty(\ln f)}\dd[2]{\vu*r}}=\frac{8\uppi Gf}{3H_0^2}\frac{\dd[3]{\rho_\mathrm{gw}}}{\dd{f}\dd[2]{\vu*r}},
    \end{equation}
    i.e., the GW energy density $\rho_\mathrm{gw}$ arriving from direction $\vu*r$ on the sky, with observed frequency in a logarithmic bin around $f$, measured in units of the cosmological critical energy density $\rho_\mathrm{c}\equiv3H_0^2/\qty(8\uppi G)$.
(Here and throughout, we use units in which $c=1$.)
We treat $\Omega\qty(f,\vu*r)$ as a random field on the sphere for each frequency $f$.
Our analysis holds for each frequency bin independently, so we drop the explicit $f$ dependence below.

Approximating the SGWB as a Gaussian random field (GRF), it is fully specified by its first two moments,
    \begin{align}
    \begin{split}
        \bar{\Omega}\qty(\vu*r)&\equiv\ev{\Omega\qty(\vu*r)}_\Omega,\\
        C\qty(\vu*r,\vu*r')&\equiv\ev{\Omega\qty(\vu*r)\Omega\qty(\vu*r')}_\Omega-\bar{\Omega}\qty(\vu*r)\bar{\Omega}\qty(\vu*r'),
    \end{split}
    \end{align}
    which we call the mean and the two-point correlation function (2PCF), respectively.
Here we have introduced a cosmological ensemble average $\ev{\cdots}_\Omega$, which we discuss further in Sec.~\ref{sec:shot}.
It is also convenient to define the spherical harmonic components (SHCs),
    \begin{equation}
    \label{eq:sgwb-shd}
        \Omega_{\ell m}\equiv\int_{S^2}\dd[2]{\vu*r}Y_{\ell m}^*\qty(\vu*r)\Omega\qty(\vu*r).
    \end{equation}
These form a set of complex random variables that fully specify any given random realisation of the field, and allow us to analyse the statistics of the SGWB on different angular scales.

For extragalactic sources, one expects the SGWB to be statistically isotropic, so that the mean is the same everywhere on the sky, and the 2PCF depends only on the angle between the two points,
    \begin{align}
    \begin{split}
        \bar{\Omega}&\equiv\ev{\Omega\qty(\vu*r)}_\Omega=\mathrm{constant},\\
        C\qty(\vu*r\vdot\vu*r')&\equiv\ev{\Omega\qty(\vu*r)\Omega\qty(\vu*r')}_\Omega-\bar{\Omega}^2.
    \end{split}
    \end{align}
An immediate consequence of the first line is that the SHCs all have zero mean, except the monopole,\footnote{%
It is important to stress the difference between the mean $\bar{\Omega}$ and the monopole $\Omega_{00}$.
Both describe the isotropic component of the SGWB, but the former is a fixed deterministic quantity that can be predicted from theory, while the latter is a random variable that must be measured.
Different cosmological observers will measure different values of the monopole, drawn from a distribution that has $\bar{\Omega}$ as its mean.
This distinction is usually not made in the case of the CMB, where fluctuations in the temperature monopole over different realisations are negligible.
However, the SGWB monopole can vary significantly between different random realisations~\cite{Meacher:2014aca}, particularly due to shot noise.
}
    \begin{equation}
    \label{eq:Omega-ell-m-mean}
        \ev{\Omega_{\ell m}}_\Omega=\delta_{\ell0}\delta_{m0}\sqrt{4\uppi}\bar{\Omega}.
    \end{equation}
(The factor of $\sqrt{4\uppi}$ here is just the integral of $Y_{00}$ over the sphere.)

The statistical properties of the anisotropies are conveniently described by a multipole expansion of the 2PCF,
    \begin{equation}
        C_\ell\equiv\int_{S^2}\dd[2]{\vu*r}P_\ell\qty(\vu*r\vdot\vu*r')C\qty(\vu*r\vdot\vu*r'),
    \end{equation}
    which we call the angular power spectrum.
Together with $\bar{\Omega}$, this completely characterises a statistically isotropic GRF, with each $\ell$ corresponding to anisotropies of angular size $\approx\uppi/\ell$.
More precisely, the contribution to the total angular variance of the SGWB per logarithmic angular scale can be approximated as
    \begin{equation}
        \dv{}{\qty(\ln\ell)}\mathrm{Var}\qty[\Omega]_\Omega\approx\frac{\ell\qty(\ell+1)}{2\uppi}C_\ell.
    \end{equation}
The $C_\ell$'s also have a very natural interpretation in terms of the SHCs, \eqref{eq:sgwb-shd}.
It is possible to show that for a statistically isotropic background,\footnote{%
    Note that here, and throughout, we define the covariance of two random variables with a complex conjugate on the second argument, i.e. $\mathrm{Cov}\qty[X,Y]\equiv\ev{XY^*}-\ev{X}\ev{Y^*}=\mathrm{Cov}\qty[Y,X]^*$.
    This ensures that the variance is always real, since $\mathrm{Var}\qty[X]\equiv\mathrm{Cov}\qty[X,X]=\ev*{\qty|X|^2}-\qty|\ev{X}|^2$.
    }
    \begin{equation}
    \label{eq:Omega-ell-m-cov}
        \mathrm{Cov}\qty[\Omega_{\ell m},\Omega_{\ell'm'}]_\Omega=\delta_{\ell\ell'}\delta_{mm'}C_\ell,
    \end{equation}
    and in particular,
    \begin{equation}
    \label{eq:Omega-ell-m-var}
        \mathrm{Var}\qty[\Omega_{\ell m}]_\Omega=C_\ell,
    \end{equation}
    i.e., each $\Omega_{\ell m}$ is uncorrelated with the others, and has variance $C_\ell$.

\subsection{The standard $C_\ell$ estimator}

We want to measure the SGWB angular power spectrum using a network of GW detectors.
The standard method for this exploits \eqref{eq:Omega-ell-m-var}, which can be rewritten as
    \begin{equation}
    \label{eq:shd-2nd-moment}
        \ev{\qty|\Omega_{\ell m}|^2}_\Omega=\qty|\ev{\Omega_{\ell m}}_\Omega|^2+C_\ell=\delta_{\ell0}\delta_{m0}4\uppi\bar{\Omega}^2+C_\ell.
    \end{equation}
We see immediately that for $\ell>0$, we can use the absolute-square of each SHC as an estimator of the corresponding multipole,
    \begin{equation}
    \label{eq:naive-estimator}
        \hat{C}^{m}_{\ell}\equiv\qty|\Omega_{\ell m}|^2.
    \end{equation}
This $m$-dependent ``naive'' estimator is unbiased, in the sense that $\ev*{\hat{C}^{m}_{\ell}}_\Omega=C_\ell$.
We can calculate its variance explicitly in the case where the SGWB is a GRF by using Isserlis' theorem to give
    \begin{equation}
        \mathrm{Var}\qty\big[\hat{C}^{m}_{\ell}]_\Omega=\qty(1+\delta_{m0})C_\ell^2.
    \end{equation}

The $m$ index runs over $-\ell\le m\le+\ell$, so in fact we have $2\ell+1$ different naive estimators, each with complementary statistical information.
We can therefore construct a better estimator by averaging over all of these.
This defines the standard $C_\ell$ estimator,
    \begin{equation}
    \label{eq:std-estimator}
        \hat{C}^{(\mathrm{std})}_\ell
        \equiv\frac{1}{2\ell+1}\sum_{m=-\ell}^{+\ell}\qty|\Omega_{\ell m}|^2.
    \end{equation}
The variance in this case is significantly lower, particularly at large $\ell$,
    \begin{equation}
    \label{eq:cosmic-var}
        \mathrm{Var}\qty\big[\hat{C}^{(\mathrm{std})}_\ell]_\Omega=\frac{2}{2\ell+1}C_\ell^2.
    \end{equation}
Note that this $1/\qty(2\ell+1)$ scaling is exactly what one should expect for $2\ell+1$ independent measurements of the same quantity.
The variance in \eqref{eq:cosmic-var} is in fact the minimum possible variance for any estimator of the angular power spectrum in the absence of shot noise, as it saturates the Cram\'er-Rao bound (see Appendix~\ref{sec:min-var}).

For the case $\ell=0$, we see from~\eqref{eq:shd-2nd-moment} that the standard estimator is biased, $\ev*{\hat{C}_0^{(\mathrm{std})}}_\Omega=4\uppi\bar{\Omega}^2+C_0$.
The problem here is that one wants to estimate the cosmological mean ($\sqrt{4\uppi}\bar{\Omega}$) and variance ($C_0$) of the observed monopole, $\Omega_{00}$, but with only a single cosmological realisation.
This is impossible, as the sample variance is undefined for a single sample; the best one can do is estimate $\bar{\Omega}$.
For this reason, the rest of this article will focus on higher multipoles, $\ell>0$.

Of course, in practice, the SHCs themselves must first be estimated from the strain data in the interferometers; this in itself is a nontrivial data analysis challenge~\cite{Romano:2016dpx}.
However, the details of this process are unimportant for the purposes of this article.

\section{Shot noise and hierarchical averaging}
\label{sec:shot}

\begin{figure}[t!]
    \includegraphics[width=0.45\textwidth]{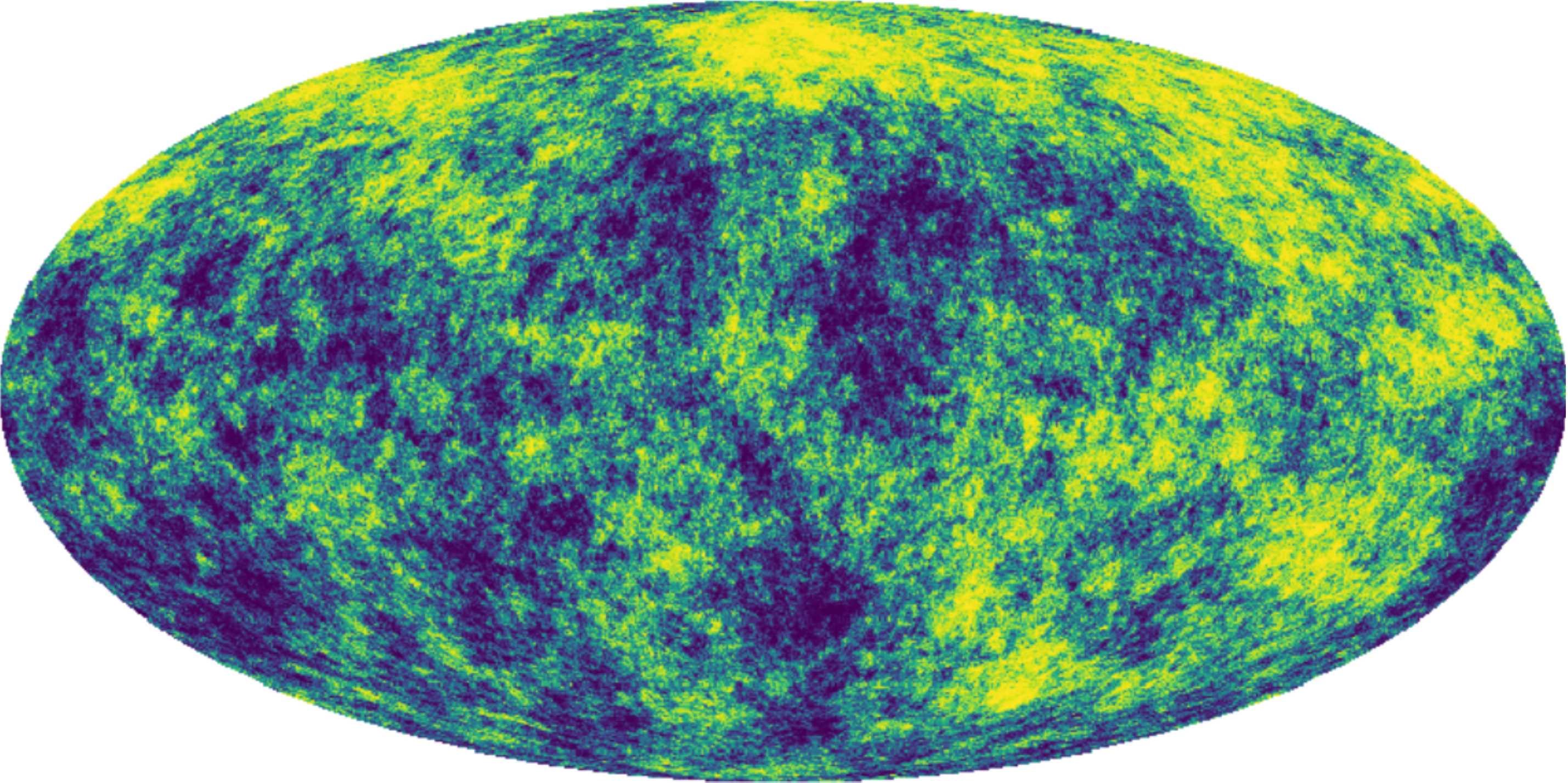}\vspace{0.05in}
    \includegraphics[width=0.45\textwidth]{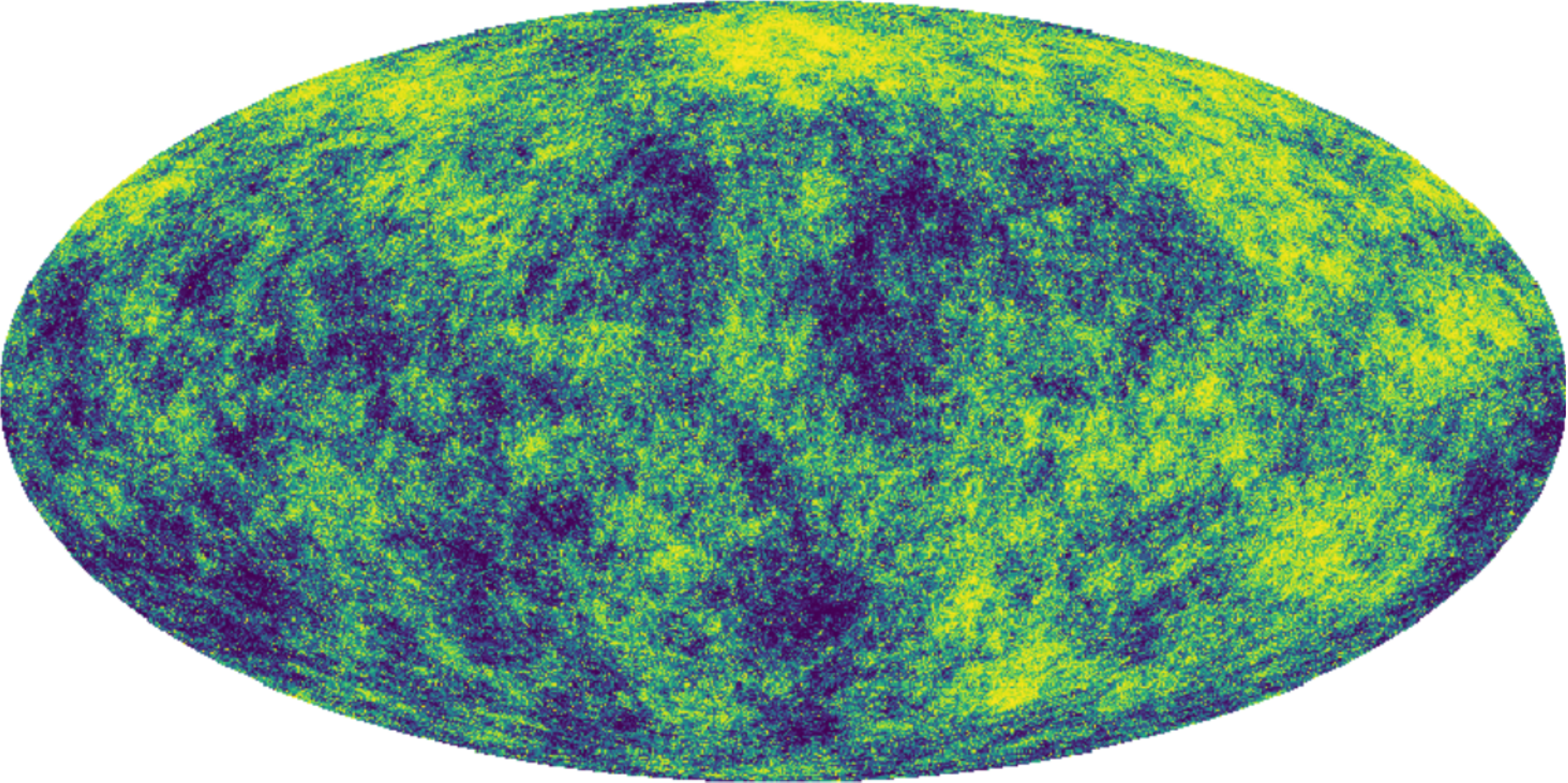}\vspace{0.05in}
    \includegraphics[width=0.45\textwidth]{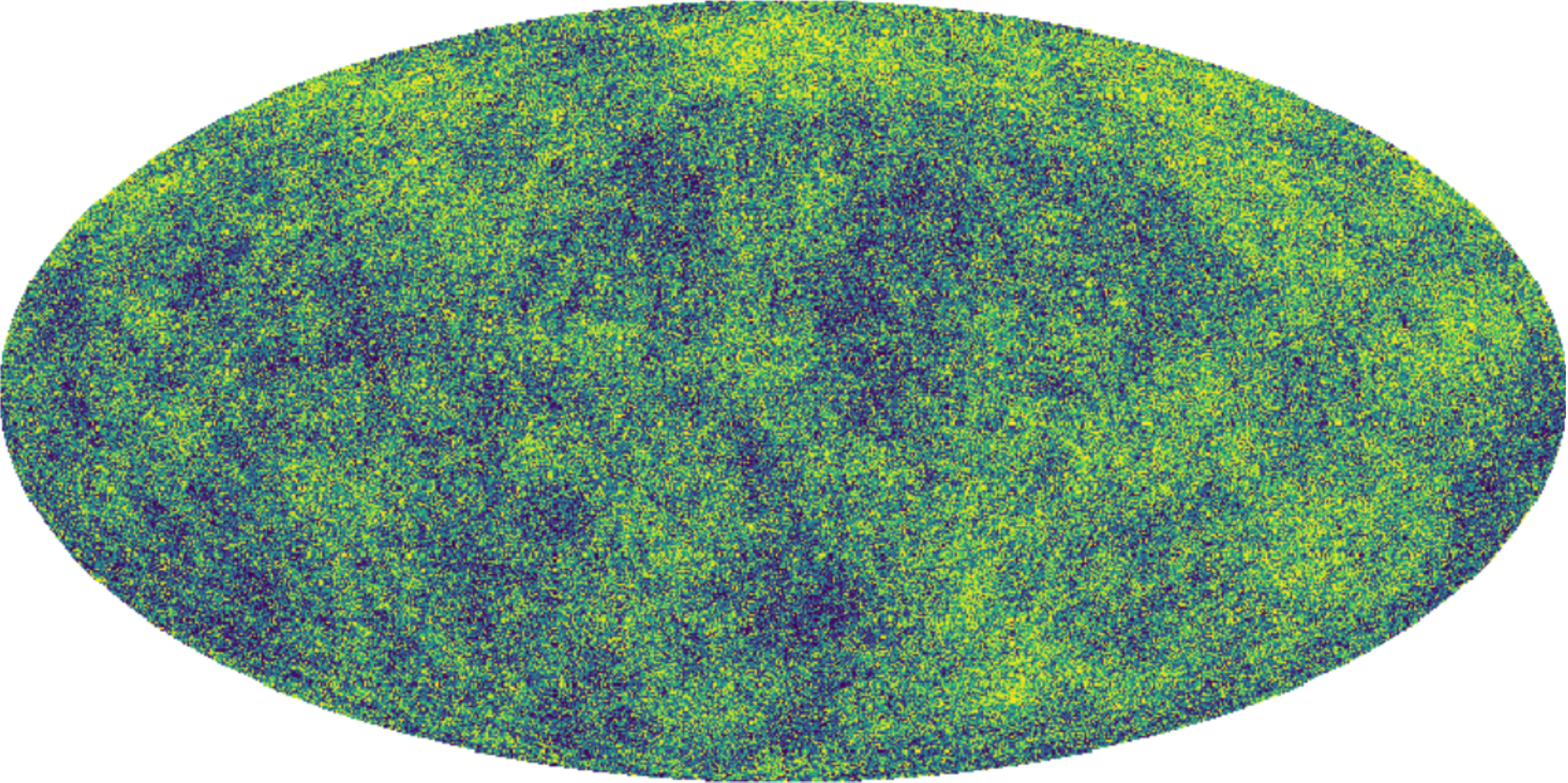}\vspace{0.05in}
    \includegraphics[width=0.45\textwidth]{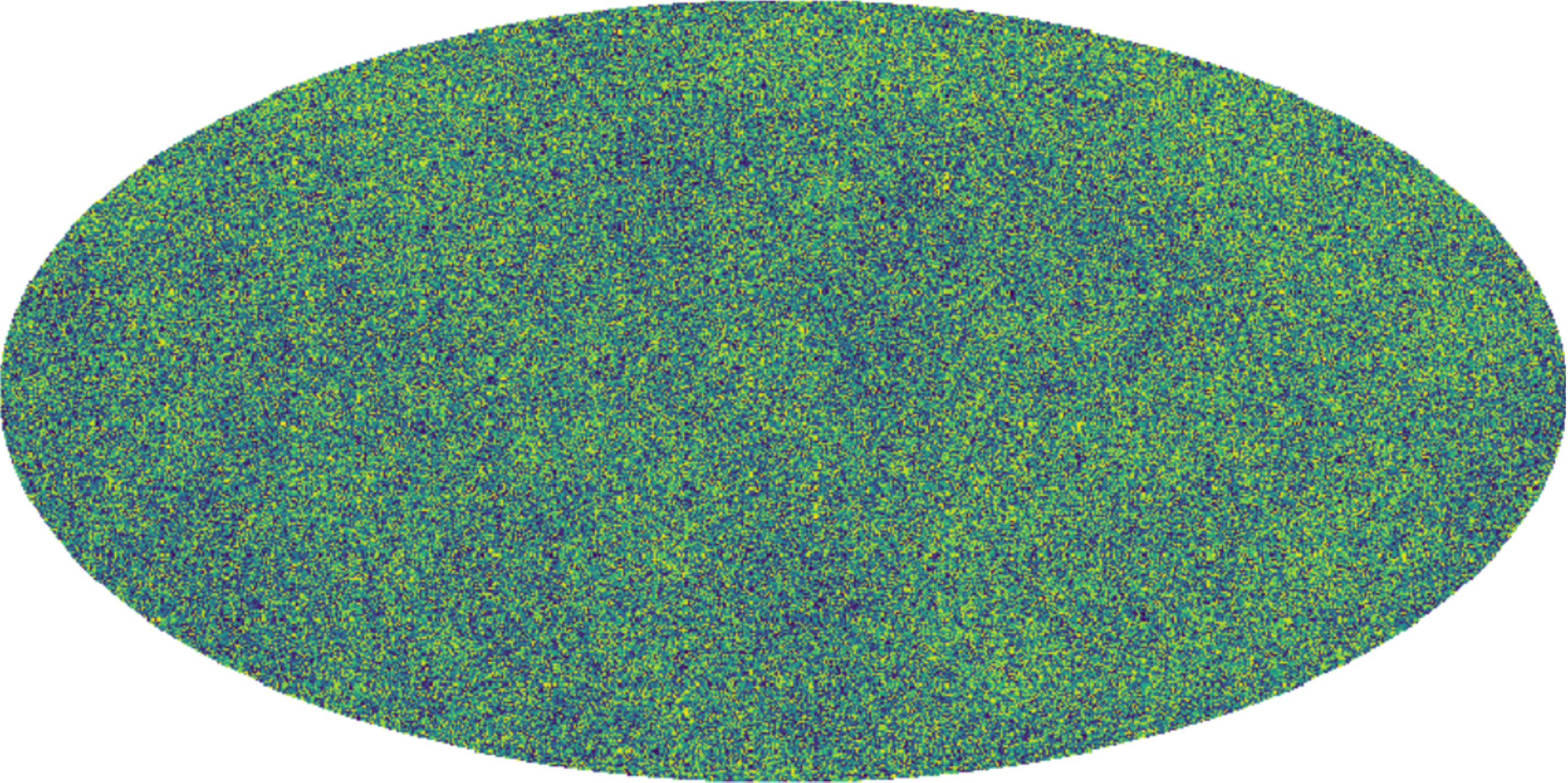}
    \caption{%
    A toy-model depiction of shot noise.
    All four images are HEALP\textsc{ix}~\cite{Gorski:2004by} maps with scale-invariant angular power spectra $\ell\qty(\ell+1)C_\ell=\mathrm{constant}$, plus varying degrees of shot noise power.
    From top to bottom, the shot noise power is equal to $\mathcal{W}=0$, $\mathcal{W}=10^{-5}\bar{\Omega}^2$, $\mathcal{W}=10^{-4}\bar{\Omega}^2$, and $\mathcal{W}=10^{-3}\bar{\Omega}^2$.
    Physically, these represent different observations of the AGWB, with different observation time intervals, leading to different levels of shot noise power.
    All four maps have the same underlying random realisation of LSS, which is why the same large-scale features can be recognised in each of them.
    However, increasing the amount of shot noise leads to much stronger anisotropies on small scales, making it harder to discern the relatively subtle large-scale features.
    }%
    \label{fig:maps}
\end{figure}

\begin{figure*}[t]
    \begin{center}
    \includegraphics[width=0.45\textwidth]{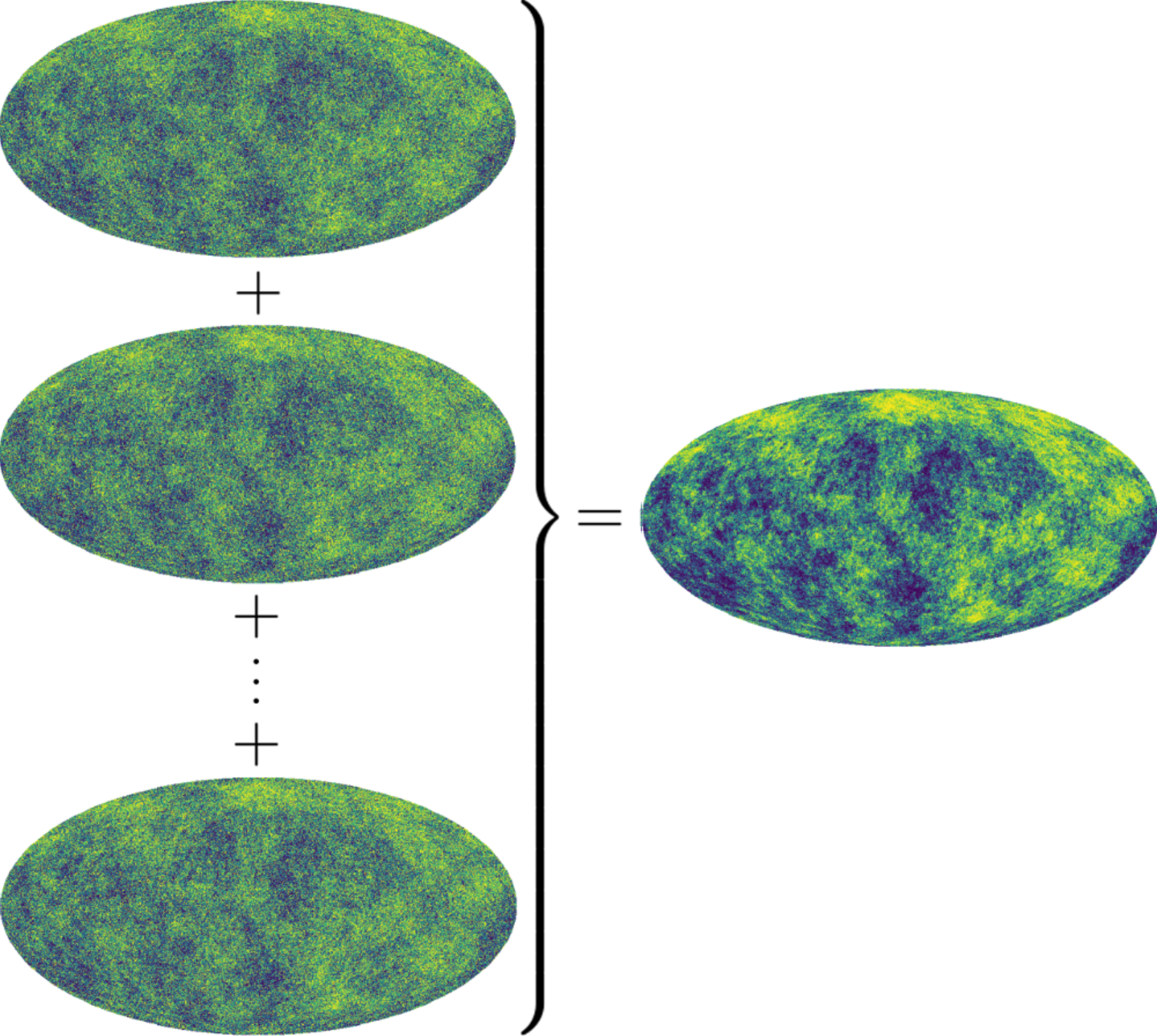}
    \hspace{0.5cm}
    \includegraphics[width=0.45\textwidth]{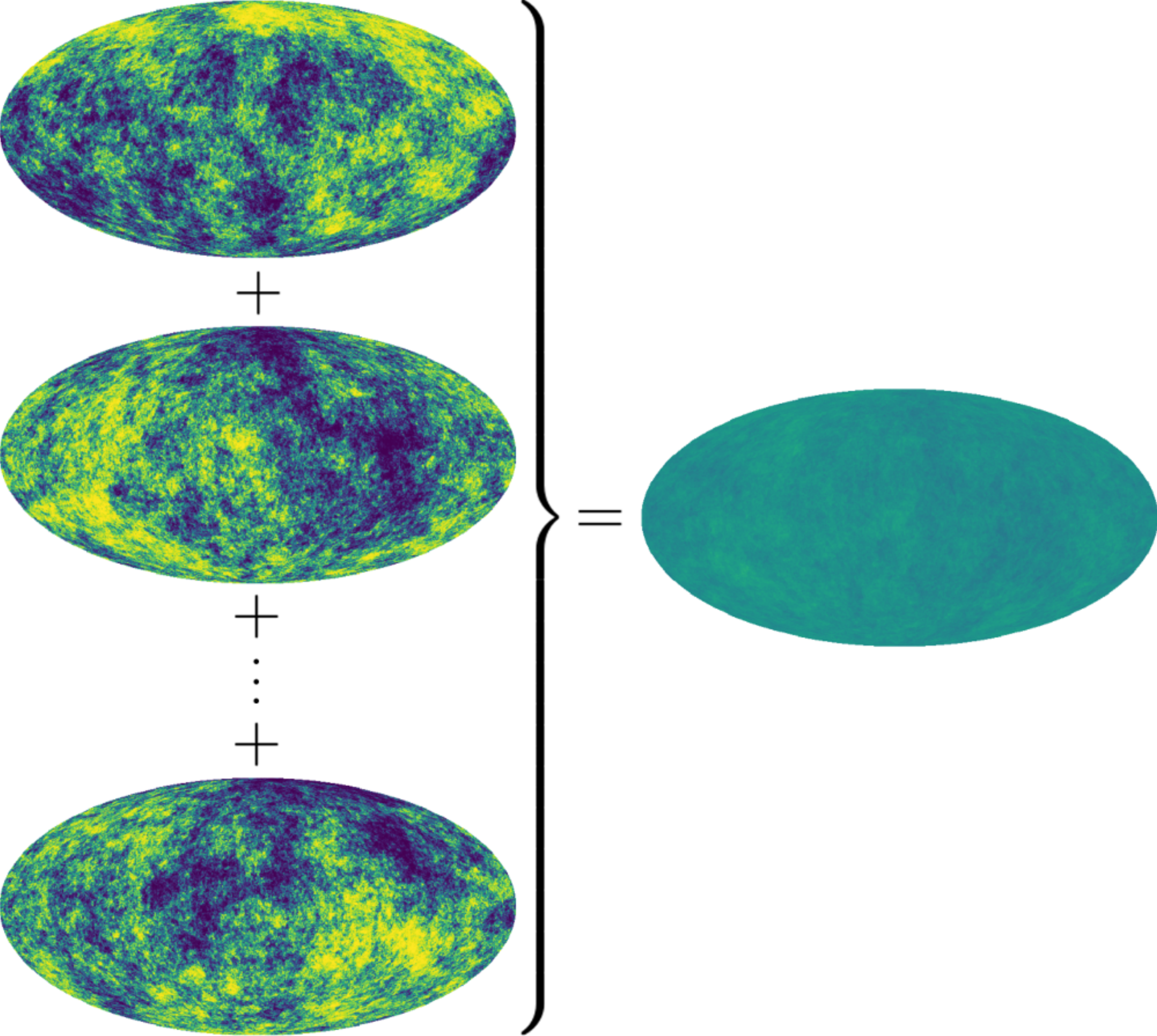}
    \end{center}
    \caption{%
    An illustration of the two averaging procedures introduced in Sec.~\ref{sec:shot}.
    The left-hand panel shows the result of averaging over ten independent realisations of shot noise (at a level of $\mathcal{W}=10^{-4}\bar{\Omega}^2$) for a single given realisation of LSS (which is the same realisation as the top map of Figure~\ref{fig:maps} and the top map in the righthand panel here).
    This is an approximation to the $\ev{\cdots}_S$ operation; in the limit of averaging over an infinite number of independent realisations, the resulting map would have zero shot noise.
    Similarly, the righthand panel shows the result of averaging over ten independent realisations of LSS, with zero shot noise.
    This is an approximation to the $\ev{\cdots}_\Omega$ operation; in the limit of averaging over an infinite number of independent realisations, the resulting map would be perfectly uniform (i.e. all SHCs equal to zero, except the monopole).
    }%
    \label{fig:avg}
\end{figure*}

\begin{figure*}[t]
    \includegraphics[width=\textwidth]{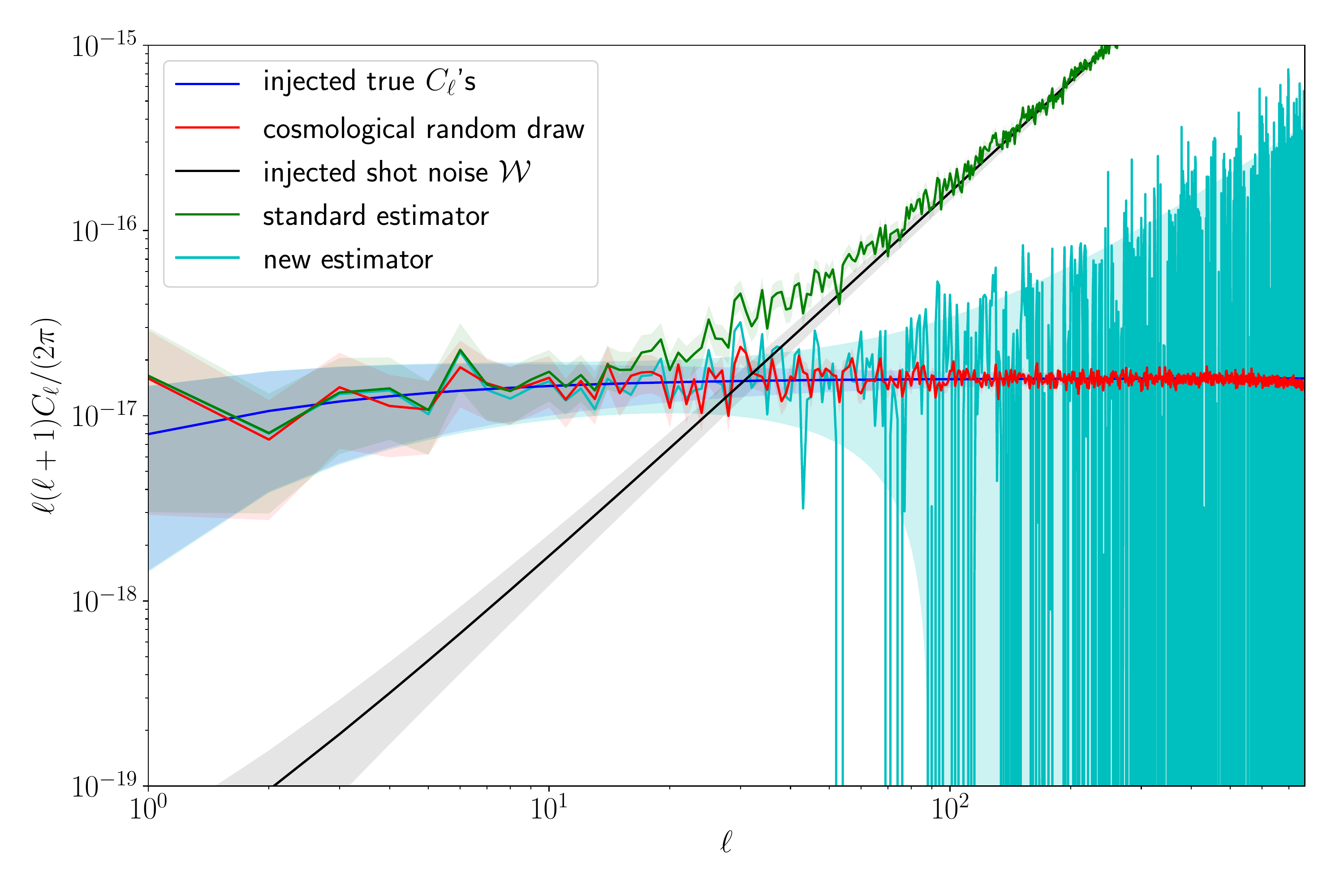}
    \caption{%
    Simulated angular power spectra using the standard estimator \eqref{eq:std-estimator} and the new estimator \eqref{eq:new-estimator}.
    The dark blue line is the chosen ``true'' spectrum to be estimated, which is here taken as scale-invariant for simplicity, $\ell\qty(\ell+1)C_\ell\approx\mathrm{constant}$.
    The red line is the spectrum of a single random cosmological realisation of the AGWB (i.e. a single Universe), distributed around the dark blue line according to cosmic variance.
    The black line is the shot-noise power $\mathcal{W}$, here set to $10^{-3}$ times the monopole.
    The green line is the spectrum resulting from the standard estimator \eqref{eq:std-estimator} for a single random realisation from the shot noise ensemble, which follows the sum of the true spectrum and the shot-noise power, $C_\ell+\mathcal{W}$.
    The cyan line is the spectrum resulting from the new estimator \eqref{eq:new-estimator}, for the same shot noise realisation, subdivided into $N_\tau=10$ independent segments.
    The shaded regions in all cases show the 1$\sigma$ uncertainty, which for the cyan line is given by \eqref{eq:var-main-result}.
    (The $C_\ell$ spectrum and shot noise power $\mathcal{W}$ used here are purely illustrative, and are not predictions for the AGWB.
    For realistic predictions of both quantities for the LIGO/Virgo frequency band, see~\cite{Jenkins:2019uzp}.)
    }%
    \label{fig:spectra}
\end{figure*}

When modelling the AGWB, one usually averages over the merger times of CBCs and the spatial locations of their host galaxies to give a smooth merger rate $R$ and galaxy number density $n$.
However, any observed realisation of the AGWB will inevitably consist of a finite number of CBCs, emitted from a finite number of galaxies.
(This was first investigated in the context of the AGWB monopole in \cite{Meacher:2014aca}.)

It was recently shown~\cite{Jenkins:2019uzp} that for a SGWB composed of a finite number of sources, the measured angular power spectrum becomes
    \begin{equation}
        C_\ell\to C_\ell+\mathcal{W},
    \end{equation}
    where $C_\ell$ is the angular power of the intrinsic, astrophysical anisotropy, and $\mathcal{W}$ represents the additional shot-noise power.
The latter is independent of $\ell$---i.e., the shot noise is spectrally white in harmonic space.
If the SGWB is composed of $N$ persistent localised sources, then schematically the shot noise scales as $\mathcal{W}\sim\bar{\Omega}^2/N$.
If the sources are transient (such as CBCs) then this becomes $\mathcal{W}_\tau\sim\bar{\Omega}^2/\qty(R\tau)$, with $R$ the event rate and $\tau$ the time interval used to measure the stochastic GW energy density.
In the latter case, we write the shot noise power as $\mathcal{W}_\tau$ to emphasise that it depends on the observer's choice of $\tau$ (subject to the condition that $\tau\gg1/f$, so that the appropriate Fourier component can be reliably measured).
Figure~\ref{fig:maps} gives a visualisation of how the shot noise power affects the GW intensity distribution on the sky.

It was shown in \cite{Jenkins:2019uzp} that, for the AGWB, the \emph{temporal} shot noise associated with finite CBCs per observation time is significantly larger than the true $C_\ell$ spectrum, while the \emph{spatial} shot noise associated with a finite galaxy number density is significantly smaller; we therefore focus exclusively on the former in this analysis.
This allows us to exploit the fact that we have observational access to multiple realisations of the temporal shot noise: for each successive observation interval $\tau$, we can obtain a set of SHCs with shot noise power $\mathcal{W}_\tau$.
The shot-noise fluctuations in each set of SHCs is associated with a different set of CBCs in a different set of galaxies, so it is immediately clear that each successive shot noise realisation is statistically independent.
This is in contrast with the spatial shot noise, for which we can only observe a single realisation (as it is tied to the random positions of galaxies).

There are two logically distinct random processes that govern the observed SHCs: the distribution of matter on large scales, and the emission of a finite number of GW signals from this matter distribution in a given observation period.
We model these processes, and their corresponding ensemble averages, in a hierarchical manner:
    \begin{enumerate}
        \item The true $\Omega_{\ell m}$ are drawn from Gaussian distributions with variance $C_\ell$.
        This process is associated with the cosmological averaging operation from Sec.~\ref{sec:angular-power-spectrum},
            \begin{equation}
                \ev{\cdots}_\Omega\equiv\text{cosmological average},
            \end{equation}
            which can be thought of as an average over an ``ensemble of Universes,'' with each Universe having a distinct random realisation of LSS.
        Of course, we only have access to a single such realisation.
        \item The true $\Omega_{\ell m}$ are modulated by shot noise, so that a set of ``noisy'' $\Omega^\mu_{\ell m}$ is drawn from a distribution with the true components $\Omega_{\ell m}$ as its mean.
        This draw is independent for each time interval, with different intervals being labelled by the index $\mu$.
        We write the associated average over shot noise realisations as
            \begin{equation}
                \ev{\cdots}_S\equiv\text{shot noise average}.
            \end{equation}
        Physically, this reflects the fact that a given cosmological distribution of galaxies can correspond to many different SGWB realisations, as the number and times-of-arrival of transient GW signals from each galaxy are essentially random.
    \end{enumerate}
Figure~\ref{fig:avg} is a schematic representation of the two averaging procedures.

More quantitatively, the first two moments of the noisy SHCs under the shot noise average for a fixed realisation of LSS are
    \begin{align}
    \begin{split}
        \ev{\Omega^\mu_{\ell m}}_S&=\Omega_{\ell m},\\
        \mathrm{Cov}\qty[\Omega^\mu_{\ell m},\Omega^\nu_{\ell'm'}]_S&=\delta_{\ell\ell'}\delta_{mm'}\delta_{\mu\nu}\mathcal{W}_\tau.
    \end{split}
    \end{align}
The first line is true by definition, while the second states that each SHC in each time interval has equal shot noise power, and is uncorrelated with all the others.\footnote{%
    The lack of correlation between time intervals is due to the acausal relationship between distant GW sources, while the lack of correlation between different $\ell m$ is due to statistical isotropy.}
Averaging also over realisations of LSS, we find
    \begin{align}
    \begin{split}
    \label{eq:cross-spectrum}
        \ev{\Omega^\mu_{\ell m}}_{S,\Omega}&=0,\\
        \mathrm{Cov}\qty[\Omega^\mu_{\ell m},\Omega^\nu_{\ell'm'}]_{S,\Omega}&=\delta_{\ell\ell'}\delta_{mm'}\qty(C_\ell+\delta_{\mu\nu}\mathcal{W}_\tau),
    \end{split}
    \end{align}
    where we have used \eqref{eq:Omega-ell-m-mean} and~\eqref{eq:Omega-ell-m-cov}, and have introduced the shorthand $\ev{\cdots}_{S,\Omega}\equiv\ev{\ev{\cdots}_S}_\Omega$.\footnote{%
    Note that $\mathrm{Cov}\qty[X,Y]_{S,\Omega}\equiv\ev{XY^*}_{S,\Omega}-\ev{X}_{S,\Omega}\ev{Y^*}_{S,\Omega}$, which is \emph{not} equal to $\ev{\mathrm{Cov}\qty[X,Y]_S}_\Omega$.}
(Recall that $\ell>0$, so the SHCs all have zero mean.)

\section{Mitigating the shot noise}
\label{sec:estimator}

In this section, we use \eqref{eq:cross-spectrum} to define a function of the noisy SHCs $\Omega_{\ell m}^\mu$ that is an unbiased estimator of the true angular power spectrum $C_\ell$ in the presence of shot noise.
We then show that (in the appropriate limit) this is the minimum-variance unbiased estimator (MVUE).

We start by modifying the standard autocorrelation estimator \eqref{eq:std-estimator}, forming a set of cross-correlations between different time intervals,
    \begin{equation}
        \hat{C}_\ell^{\mu\nu}\equiv\frac{1}{2\ell+1}\sum_{m=-\ell}^{+\ell}\Omega_{\ell m}^\mu\Omega_{\ell m}^{\nu*}.
    \end{equation}
These are unbiased if and only if $\mu\ne\nu$,
    \begin{equation}
        \ev*{\hat{C}_\ell^{\mu\nu}}_{S,\Omega}=C_\ell+\delta_{\mu\nu}\mathcal{W}_\tau.
    \end{equation}
We can combine these estimators in much the same way that we combine multiple ``naive'' estimators \eqref{eq:naive-estimator} to form the standard estimator \eqref{eq:std-estimator}.
Suppose that our total observing time is $T$, so that there are $N_\tau\equiv T/\tau$ time segments.
Then we have $N_\tau\qty(N_\tau-1)/2$ pairs $\mu,\nu$ for which $\mu\ne\nu$.
Summing over these, we define the combined estimator
    \begin{equation}
    \label{eq:new-estimator}
        \hat{C}_\ell\equiv\frac{2}{N_\tau\qty(N_\tau-1)}\sum_{\mu=1}^{N_\tau}\sum_{\nu=\mu+1}^{N_\tau}\hat{C}_\ell^{\mu\nu}.
    \end{equation}
This is unbiased, $\ev*{\hat{C}_\ell}_{S,\Omega}=C_\ell$, with variance given by
\begin{widetext}
    \begin{equation}
    \label{eq:var-4th-moment}
        \mathrm{Var}\big[\hat{C}_\ell\big]_{S,\Omega}=\qty[\frac{2}{N_\tau\qty(N_\tau-1)\qty(2\ell+1)}]^2\sum_{m=-\ell}^{+\ell}\sum_{m'=-\ell}^{+\ell}\sum_{\mu=1}^{N_\tau}\sum_{\mu'=1}^{N_\tau}\sum_{\nu=\mu+1}^{N_\tau}\sum_{\nu'=\mu'+1}^{N_\tau}\mathrm{Cov}\qty[\Omega_{\ell m}^\mu\Omega_{\ell m}^{\nu*},\Omega_{\ell m'}^{\mu'}\Omega_{\ell m'}^{\nu'*}]_{S,\Omega}.
    \end{equation}
\end{widetext}

Evaluating \eqref{eq:var-4th-moment} requires us to evaluate the fourth moment of the noisy SHCs.
This would be trivial if the SHCs were all Gaussian, but we must account for the Poisson-like nature of the shot noise.
In Appendix~\ref{sec:4th-moment} we calculate the fourth moment using the same statistical model for the CBC rate density as in \cite{Jenkins:2019uzp}; this results in
    \begin{equation}
    \label{eq:var-main-result}
        \mathrm{Var}\big[\hat{C}_\ell\big]_{S,\Omega}=\frac{2}{2\ell+1}\qty[C_\ell^2+\frac{2\mathcal{W}_\tau C_\ell}{N_\tau}+\frac{\mathcal{W}_\tau^2}{N_\tau\qty(N_\tau-1)}].
    \end{equation}
This expression is tied to the fact that we have excluded the on-diagonal terms $\mu=\nu$ when constructing \eqref{eq:new-estimator}; otherwise, there would be additional contributions to the variance (the estimator would also no longer be unbiased).
Note that since $\mathcal{W}_\tau\propto1/\tau$, we have $\mathcal{W}_\tau/N_\tau\propto1/T$.
This means that we can't ``win'' by decreasing the length of the data segments $\tau$, only by increasing the total observing time $T$.
In fact, writing $\mathcal{W}_T=\mathcal{W}_\tau\qty(\tau/T)=\mathcal{W}_\tau/N_\tau$, we see that in the limit where $N_\tau\gg1$, \eqref{eq:var-main-result} becomes
    \begin{equation}
    \label{eq:var-limit}
        \mathrm{Var}\big[\hat{C}_\ell\big]_{S,\Omega}\simeq\frac{2}{2\ell+1}\qty(C_\ell+\mathcal{W}_T)^2.
    \end{equation}
This is exactly the standard cosmic variance expression from \eqref{eq:cosmic-var}, but with $C_\ell$ replaced by $C_\ell+\mathcal{W}_T$.
In Appendix~\ref{sec:min-var}, we show that this is in fact the minimum possible variance of any unbiased estimator for the $C_\ell$'s in the presence of shot noise, saturating the Cram\'er-Rao bound~\cite{Kay:1993}.
The estimator \eqref{eq:new-estimator} is therefore the MVUE in the limit $N_\tau\gg1$.

At the opposite extreme, for the minimum number of segments, $N_\tau=2$, the variance is nearly twice as large (taking $\mathcal{W}_T\gg C_\ell$),
    \begin{equation}
        \mathrm{Var}\big[\hat{C}_\ell\big]_{S,\Omega}\approx\frac{4}{2\ell+1}\mathcal{W}_T^2.
    \end{equation}
An illustrative example for $N_\tau=10$ is shown in Fig.~\ref{fig:spectra}.

The term $\mathcal{W}_T$ in \eqref{eq:var-limit} is the same as that appearing in the mean of the standard estimator, $\ev*{C_\ell^{(\mathrm{std})}}_{S,\Omega}=C_\ell+\mathcal{W}_T$, so this new optimal estimator is still affected by the presence of shot noise; the crucial improvement is that the shot noise only adds to the variance of the estimator, and does not bias the spectrum as in the standard case.

\section{Conclusion}

We have developed a new method for estimating the true, astrophysical angular power spectrum of a statistically-isotropic AGWB, in a way that is not biased by the presence of temporal shot noise.
The estimator \eqref{eq:new-estimator} is based on the simple principle of cross-correlating between statistically-independent time intervals.
It is straightforward to compute from LIGO/Virgo strain data using existing pipelines, and in the limit of a large number of time intervals (regardless of their length), it achieves the lowest possible variance of any estimator on the same data.
This is a very significant improvement upon the standard analysis, given that the shot noise power is expected to exceed the astrophysical angular power by a factor of $\approx10^4$~\cite{Jenkins:2019uzp}.
We expect that these results will be highly valuable for future GW observing campaigns, and may give us access to interesting and novel cosmological information.

\begin{acknowledgments}
    We thank Andrew Matas for reading the manuscript and providing us with useful feedback.
    A.C.J. thanks Eugene Lim for valuable discussions.
    This article has been assigned document number LIGO-P1900209.
    Some of the results in this article have been derived using the HEALP\textsc{ix} package~\cite{Gorski:2004by}.
    A.C.J. is supported by King's College London through a Graduate Teaching Scholarship.
    J.D.R. acknowledges support from National Science Foundation Award No. PHY-1505861
    and start-up funds from Texas Tech University.
    M.S. is supported in part by the Science and Technology Facility Council (STFC), United Kingdom, under the research grant No. ST/P000258/1.
\end{acknowledgments}

\appendix
\section{Fourth moment of the spherical harmonic components with shot noise}
\label{sec:4th-moment}

Here we compute the fourth moment of the noisy SHCs,
    \begin{equation}
        \ev{\Omega_{\ell m}^\mu\Omega_{\ell m}^{\nu*}\Omega_{\ell m'}^{\mu'*}\Omega_{\ell m'}^{\nu'}}_{S,\Omega},
    \end{equation}
using the same statistical model for the AGWB as in \cite{Jenkins:2019uzp}.

In a given volume element $\updelta V$, the number of galaxies is modelled as a Poisson random variable $N\sim\mathrm{Pois}\qty[\bar{n}\updelta V]$, where $\bar{n}$ is the mean galaxy number density,\footnote{%
    A more sophisticated approach would account for the statistical properties of halos~\cite{Cooray:2002dia}.
    However, this simple assumption is sufficient for the calculation here.}
    and with the number of CBCs in each galaxy forming a set of independent and identically distributed Poisson random variables $\lambda_i\sim\mathrm{Pois}\qty[R\tau_\mathrm{s}]$, where $R$ is the mean rate per galaxy, and $\tau_\mathrm{s}=\tau/\qty(1+z)$ is the source-frame time interval of the observation.
The total CBC count in the volume element is then
    \begin{equation}
        \Lambda\equiv\sum_{i=1}^N\lambda_i,
    \end{equation}
    which follows a compound Poisson distribution, with mean
    \begin{equation}
        \ev{\Lambda}_S=\ev{\lambda}_S\ev{N}_S=R\tau_\mathrm{s}\bar{n}\updelta V\equiv\bar{\Lambda}.
    \end{equation}
Here we characterise this distribution by its cumulants,
    \begin{equation}
        \kappa_n\equiv\left.\dv[n]{K\qty(x)}{x}\right|_{x=0},
    \end{equation}
    where $K\qty(x)$ is the cumulant-generating function,
    \begin{align}
    \begin{split}
        K\qty(x)&\equiv\ln\ev{\mathrm{e}^{x\Lambda}}_S\\
        &=\ev{N}_S\qty[\mathrm{e}^{\ev{\lambda}_S\qty(\mathrm{e}^x-1)}-1]\\
        &=\bar{n}\updelta V\qty[\mathrm{e}^{R\tau_\mathrm{s}\qty(\mathrm{e}^x-1)}-1].
    \end{split}
    \end{align}
The first two cumulants, $\kappa_1,\kappa_2$, are just the mean and the variance.
Higher cumulants $\kappa_n$ represent the ``connected components'' of the $n$th moments of the distribution, and vanish if the distribution is Gaussian.
The fourth moment of $\Lambda$ is given in terms of cumulants as
    \begin{equation}
        \ev{\Lambda^4}_S=\kappa_1^4+6\kappa_2\kappa_1^2+3\kappa_2^2+4\kappa_3\kappa_1+\kappa_4.
    \end{equation}

Consider now a set of different volume elements observed at different times $\updelta V_\mu$, whose total CBC counts $\Lambda_\mu$ are independent and identically distributed.
Their fourth moment is then
    \begin{align}
    \begin{split}
        \ev{\Lambda_\mu\Lambda_\nu\Lambda_\rho\Lambda_\sigma}_S&=\kappa_1^4\\
        &+\kappa_2\kappa_1^2\qty(\delta_{\mu\nu}+\delta_{\mu\rho}+\delta_{\mu\sigma}+\delta_{\nu\rho}+\delta_{\nu\sigma}+\delta_{\rho\sigma})\\
        &+\kappa_2^2\qty(\delta_{\mu\nu}\delta_{\rho\sigma}+\delta_{\mu\rho}\delta_{\nu\sigma}+\delta_{\mu\sigma}\delta_{\nu\rho})\\
        &+\kappa_3\kappa_1\qty(\delta_{\mu\nu\rho}+\delta_{\mu\nu\sigma}+\delta_{\mu\rho\sigma}+\delta_{\nu\rho\sigma})\\
        &+\kappa_4\delta_{\mu\nu\rho\sigma},
    \end{split}
    \end{align}
    where $\delta_{\mu\nu\rho}$ is non-zero only if $\mu=\nu=\rho$, etc.
The first four cumulants are
    \begin{align}
    \begin{split}
        \kappa_1&=\bar{\Lambda},\\
        \kappa_2&=\bar{\Lambda}\qty[1+R\tau_\mathrm{s}],\\
        \kappa_3&=\bar{\Lambda}\qty[1+3R\tau_\mathrm{s}+(R\tau_\mathrm{s})^2],\\
        \kappa_4&=\bar{\Lambda}\qty[1+7R\tau_\mathrm{s}+6(R\tau_\mathrm{s})^2+(R\tau_\mathrm{s})^3].
    \end{split}
    \end{align}
The rate per galaxy is typically on the order of $\mathrm{Myr}^{-1}$, while the source-frame observing time is at most a few years, so we we have $R\tau_\mathrm{s}\sim10^{-6}$.
So to a very good approximation, we can take
    \begin{equation}
        \kappa_1\approx\kappa_2\approx\kappa_3\approx\kappa_4\approx\bar{\Lambda}.
    \end{equation}

In order to compute the fourth moment of the SHCs $\Omega_{\ell m}^\mu$, we translate from the CBC number count to the comoving CBC rate density,
    \begin{equation}
        \mathcal{R}\qty(\vb*r,\vb*\zeta)\equiv nR=\lim_{\updelta V\to0}\frac{\Lambda}{\tau_\mathrm{s}\updelta V},
    \end{equation}
    and replace $\bar{\Lambda}$ with $\bar{\mathcal{R}}\equiv\bar{n}R$.
We introduce $\vb*\zeta$ to represent the parameters of the CBC (masses, spins, \dots) and of the galaxy (star formation rate, metallicity, \dots).
Making this replacement, we obtain
    \begin{align}
    \begin{split}
    \label{eq:R-4pt}
        \ev{\mathcal{R}_\mu\mathcal{R}_\nu\mathcal{R}_\rho\mathcal{R}_\sigma}_S&=\bar{\mathcal{R}}_\mu\bar{\mathcal{R}}_\nu\bar{\mathcal{R}}_\rho\bar{\mathcal{R}}_\sigma\\
        &+\bar{\mathcal{R}}_\mu\bar{\mathcal{R}}_\nu\bar{\mathcal{R}}_\rho\frac{\delta_{\rho\sigma}}{\tau_{s,\rho}}+5\,\mathrm{perms.}\\
        &+\bar{\mathcal{R}}_\mu\bar{\mathcal{R}}_\rho\frac{\delta_{\mu\nu}}{\tau_{s,\mu}}\frac{\delta_{\rho\sigma}}{\tau_{s,\rho}}+2\,\mathrm{perms.}\\
        &+\bar{\mathcal{R}}_\mu\bar{\mathcal{R}}_\nu\frac{\delta_{\nu\rho\sigma}}{\tau_{s,\nu}^2}+3\,\mathrm{perms.}\\
        &+\bar{\mathcal{R}}_\mu\frac{\delta_{\mu\nu\rho\sigma}}{\tau_{s,\mu}^3},
    \end{split}
    \end{align}
    where here the delta symbols are shorthand for
    \begin{equation}
    \label{eq:delta-shorthand}
        \delta_{\mu\nu}\to\delta_{\mu\nu}\delta^3\qty(\vb*r_\mu-\vb*r_\nu)\delta\qty(\vb*\zeta_\mu,\vb*\zeta_\nu),
    \end{equation}
    i.e., two GW sources must coincide in space, in time, and in parameter space, in order to contribute to the shot noise.

Now, using the equation for the AGWB in terms of the CBC rate density from \cite{Jenkins:2019uzp}, we write
    \begin{align}
    \begin{split}
        &\ev{\Omega_{\ell m}^\mu\Omega_{\ell m}^{\nu*}\Omega_{\ell m'}^{\mu'*}\Omega_{\ell m'}^{\nu'}}_S\\
        &=\qty(\frac{\uppi}{3})^4\qty(t_H\nu_\mathrm{o})^{12}\int\frac{\dd[3]{\vb*r_\mu}\dd[3]{\vb*r_\nu}\dd[3]{\vb*r_{\mu'}}\dd[3]{\vb*r_{\nu'}}}{r_H^{12}}Y_{\ell m}^{\mu*}Y_{\ell m}^\nu Y_{\ell m'}^{\mu'}Y_{\ell m'}^{\nu'*}\\
        &\times\qty(1+z_\mu)\qty(1+z_\nu)\qty(1+z_{\mu'})\qty(1+z_{\nu'})\qty(\frac{r_H^4}{r_\mu r_\nu r_{\mu'}r_{\nu'}})^2\\
        &\times\int\dd{\vb*\zeta_\mu}\dd{\vb*\zeta_\nu}\dd{\vb*\zeta_{\mu'}}\dd{\vb*\zeta_{\nu'}}\mathcal{S}_\mu\mathcal{S}_\nu\mathcal{S}_{\mu'}\mathcal{S}_{\nu'}\ev{\mathcal{R}_\mu\mathcal{R}_\nu\mathcal{R}_{\mu'}\mathcal{R}_{\nu'}}_S,
    \end{split}
    \end{align}
    where $Y_{\ell m}^\mu$ is shorthand for $Y_{\ell m}\qty(\vu*r_\mu)$, etc., and $\mathcal{S}\qty(\nu_\mathrm{s},\vb*\zeta)$ encodes the GW emission of each CBC as a function of source-frame frequency $\nu_\mathrm{s}$, and has dimensions of $\qty[\mathrm{time}]^4$ in units where $c=1$.
Using \eqref{eq:R-4pt} and the properties of the spherical harmonics, this becomes
    \begin{align}
    \begin{split}
    \label{eq:Omega-4pt}
        &\ev{\Omega_{\ell m}^\mu\Omega_{\ell m}^{\nu*}\Omega_{\ell m'}^{\mu'*}\Omega_{\ell m'}^{\nu'}}_S\\
        &\quad=\qty|\Omega_{\ell m}|^2\qty|\Omega_{\ell m'}|^2\\
        &\quad+\mathcal{W}_\tau\qty|\Omega_{\ell m}|^2\qty[\delta_{\mu'\nu'}+\delta_{m,-m'}\qty(\delta_{\nu\mu'}+\delta_{\mu\nu'})]\\
        &\quad+\mathcal{W}_\tau\qty|\Omega_{\ell m'}|^2\qty[\delta_{\mu\nu}+\delta_{mm'}\qty(\delta_{\mu\mu'}+\delta_{\nu\nu'})]\\
        &\quad+\mathcal{W}^2_\tau\qty(\delta_{\mu\nu}\delta_{\mu'\nu'}+\delta_{mm'}\delta_{\mu\mu'}\delta_{\nu\nu'}+\delta_{m,-m'}\delta_{\mu\nu'}\delta_{\nu\mu'})\\
        &\quad+c^{(3)}_{\ell m}\mathcal{X}_\tau\Omega_{\ell0}\delta_{m'0}\qty(\delta_{\mu\nu\mu'}+\delta_{\mu\nu\nu'})\\
        &\quad+c^{(3)}_{\ell m'}\mathcal{X}_\tau\Omega_{\ell0}\delta_{m0}\qty(\delta_{\mu\mu'\nu'}+\delta_{\nu\mu'\nu'})\\
        &\quad+c^{(4)}_{\ell m m'}\mathcal{Y}_\tau\delta_{\mu\nu\mu'\nu'},
    \end{split}
    \end{align}
    where the shorthand~\eqref{eq:delta-shorthand} is now no longer used.
The coefficients on the 3- and 4-point terms are defined as
    \begin{align}
    \begin{split}
        c^{(3)}_{\ell m}&\equiv\qty(-1)^m\frac{\qty(2\ell+1)^{3/2}}{\sqrt{4\uppi}}
        \begin{pmatrix}
            \ell & \ell & \ell \\
            0 & 0 & 0
        \end{pmatrix}
        \begin{pmatrix}
            \ell & \ell & \ell \\
            0 & m & -m
        \end{pmatrix},\\
        c^{(4)}_{\ell m m'}&\equiv\sum_{\ell'=0}^{2\ell}\qty(-1)^{m+m'}\frac{\qty(2\ell'+1)\qty(2\ell+1)^2}{4\uppi}\\
        &\qquad\times
        \begin{pmatrix}
            \ell & \ell & \ell' \\
            -m & m & 0
        \end{pmatrix}
        \begin{pmatrix}
            \ell & \ell & \ell' \\
            m' & -m' & 0
        \end{pmatrix}
        \begin{pmatrix}
            \ell & \ell & \ell' \\
            0 & 0 & 0
        \end{pmatrix}^2,
    \end{split}
    \end{align}
    and are just combinations of Wigner $3j$ symbols.
In addition to the 2-point term $\mathcal{W}_\tau$ we defined before, we now have 3-point and 4-point terms that appear,
    \begin{align}
    \begin{split}
        \mathcal{W}_\tau&\equiv\qty(\frac{\uppi}{3})^2\frac{\qty(t_H\nu_\mathrm{o})^6}{\tau}\int\dd{r}\frac{\qty(1+z)^3}{r_H^2r^2}\int\dd{\vb*\zeta}\bar{\mathcal{R}}\mathcal{S}^2,\\
        \mathcal{X}_\tau&\equiv\qty(\frac{\uppi}{3})^3\frac{\qty(t_H\nu_\mathrm{o})^9}{\tau^2}\int\dd{r}\frac{\qty(1+z)^5}{r_H^3r^4}\int\dd{\vb*\zeta}\bar{\mathcal{R}}\mathcal{S}^3,\\
        \mathcal{Y}_\tau&\equiv\qty(\frac{\uppi}{3})^4\frac{\qty(t_H\nu_\mathrm{o})^{12}}{\tau^3}\int\dd{r}\frac{\qty(1+z)^7}{r_H^4r^6}\int\dd{\vb*\zeta}\bar{\mathcal{R}}\mathcal{S}^4.
    \end{split}
    \end{align}

Taking the cosmological average of \eqref{eq:Omega-4pt} and subtracting the 2nd moments, we therefore find
    \begin{align}
    \begin{split}
        &\mathrm{Cov}\qty[\Omega_{\ell m}^\mu\Omega_{\ell m}^{\nu*},\Omega_{\ell m'}^{\mu'}\Omega_{\ell m'}^{\nu'*}]_{S,\Omega}\\
        &\quad=\delta_{mm'}\qty(C_\ell+\delta_{\mu\mu'}\mathcal{W}_\tau)\qty(C_\ell+\delta_{\nu\nu'}\mathcal{W}_\tau)\\
        &\quad+\delta_{m,-m'}\qty(C_\ell+\delta_{\mu\nu'}\mathcal{W}_\tau)\qty(C_\ell+\delta_{\nu\mu'}\mathcal{W}_\tau)\\
        &\quad+c^{(4)}_{\ell mm'}\mathcal{Y}_\tau\delta_{\mu\nu\mu'\nu'}.
    \end{split}
    \end{align}
Note that the term proportional to $\mathcal{X}_\tau$ has vanished, due to the SHCs having zero mean for $\ell>0$ (recall that we do not consider $\ell=0$ here).
The term proportional to $\mathcal{Y}_\tau$ is associated with the fourth cumulant $\kappa_4$, and is therefore a sign of the non-Gaussian nature of the shot noise fluctuations (as $\kappa_n=0$ for $n>2$ in the Gaussian case).
However, this term does not end up contributing to the variance of the estimator $\hat{C}_\ell$, as this includes only off-diagonal pairs $\mu\ne\nu$ and $\mu'\ne\nu'$ by design.

\section{Minimum-variance estimation of the angular power spectrum}
\label{sec:min-var}

Here we show that the estimator in \eqref{eq:new-estimator} is the MVUE for the true angular power spectrum $C_\ell$ in the presence of shot noise.
To do so, we use the fact that the variance of any estimator $\hat{\theta}\qty(x)$ for a parameter $\theta$ of a probability distribution $p\qty(x|\theta)$ obeys the Cram\'er-Rao bound~\cite{Kay:1993},
    \begin{equation}
        \mathrm{Var}[\hat{\theta}]\ge-\ev{\pdv[2]{\mathcal{L}}{\theta}}^{-1},
    \end{equation}
    where $\mathcal{L}\qty(x|\theta)\equiv\ln p\qty(x|\theta)$ is the log-likelihood.

We start by considering the case with no shot noise.
Assuming Gaussianity, the joint log-likelihood for the SHCs is then
    \begin{equation}
        \mathcal{L}=-\frac{1}{2}\sum_{\ell>0}^{\infty}\sum_{m=-\ell}^{+\ell}\qty[\ln(2\uppi C_\ell)+\frac{\qty|\Omega_{\ell m}|^2}{C_\ell}],
    \end{equation}
    and we find that
    \begin{align}
    \begin{split}
        \mathrm{Var}\big[\hat{C}_\ell\big]_\Omega&\ge\ev{\sum_{m=-\ell}^{+\ell}\frac{\qty|\Omega_{\ell m}|^2}{C_\ell^3}-\frac{1}{2C_\ell^2}}_\Omega^{-1}\\
        &=\qty(\sum_{m=-\ell}^{+\ell}\frac{1}{2C_\ell^2})^{-1}\\
        &=\frac{2}{2\ell+1}\,C_\ell^2\,.
    \end{split}
    \end{align}
This shows that the standard cosmic variance expression \eqref{eq:cosmic-var} is the best one can do in the zero-shot-noise case, as it saturates the Cram\'er-Rao bound.

Now we include shot noise, and consider the noisy SHCs, $\Omega_{\ell m}^\mu$.
Though we know these are not Gaussian (see Appendix~\ref{sec:4th-moment}), the Gaussian case is by far the most tractable, so we consider it first.
The joint Gaussian log-likelihood is fully specified by \eqref{eq:cross-spectrum},
    \begin{align}
    \label{eq:log-likelihood}
        \mathcal{L}=-\frac{1}{2}\sum_{\ell>0}^\infty\sum_{m=-\ell}^{+\ell}\qty[\ln(\mathrm{det}\qty(2\uppi\mathsf{C}_\ell))+\vb*\Omega_{\ell m}^\dagger\mathsf{C}_\ell^{-1}\vb*\Omega_{\ell m}],
    \end{align}
    where $\vb*\Omega_{\ell m}=(\Omega^1_{\ell m},\ldots,\Omega^{N_\tau}_{\ell m})$ is a vector of the noisy SHCs for a given $\ell m$, and $\mathsf{C}_\ell$ is the corresponding $N_\tau\times N_\tau$ covariance matrix,
    \begin{equation}
        \mathsf{C}_\ell=
        \begin{pmatrix}
            C_\ell+\mathcal{W}_\tau & C_\ell & \cdots & C_\ell \\
            C_\ell & C_\ell+\mathcal{W}_\tau & \cdots & C_\ell \\
            \vdots & \vdots & \ddots & \vdots \\
            C_\ell & C_\ell & \cdots & C_\ell+\mathcal{W}_\tau
        \end{pmatrix}.
    \end{equation}
One can show that this has determinant
    \begin{equation}
        \mathrm{det}\mathsf{C}_\ell=\qty(N_\tau C_\ell+\mathcal{W}_\tau)\mathcal{W}_\tau^{N_\tau-1},
    \end{equation}
    and inverse
\begin{widetext}
    \begin{equation}
        \mathsf{C}_\ell^{-1}=\frac{1}{\qty(N_\tau C_\ell+\mathcal{W}_\tau)\mathcal{W}_\tau}
        \begin{pmatrix}
            \qty(N_\tau-1)C_\ell+\mathcal{W}_\tau & -C_\ell & \cdots & -C_\ell \\
            -C_\ell & \qty(N_\tau-1)C_\ell+\mathcal{W}_\tau & \cdots & -C_\ell \\
            \vdots & \vdots & \ddots & \vdots \\
            -C_\ell & -C_\ell & \cdots & \qty(N_\tau-1)C_\ell+\mathcal{W}_\tau
        \end{pmatrix},
    \end{equation}
    using the matrix determinant lemma and the Sherman-Morrison formula, respectively.
Taking the second derivative with respect to $C_\ell$, we find
    \begin{align}
    \begin{split}
    \label{eq:cramer-rao-shot-noise}
        \mathrm{Var}\qty[\hat{C}_\ell]_{S,\Omega}&\ge-\ev{\pdv[2]{\mathcal{L}}{C_\ell}}^{-1}_{S,\Omega}\\
        &=-\ev{\sum_{m=-\ell}^{+\ell}\frac{1}{2\qty(C_\ell+\mathcal{W}_T)^2}-\frac{1}{N_\tau^2\qty(C_\ell+\mathcal{W}_T)^3}\sum_{\mu=1}^{N_\tau}\sum_{\nu=1}^{N_\tau}\Omega^{\mu*}_{\ell m}\Omega^{\nu}_{\ell m}}^{-1}_{S,\Omega}\\
        &=\qty[\sum_{m=-\ell}^{+\ell}-\frac{1}{2\qty(C_\ell+\mathcal{W}_T)^2}+\frac{1}{N_\tau^2\qty(C_\ell+\mathcal{W}_T)^3}\sum_{\mu=1}^{N_\tau}\sum_{\nu=1}^{N_\tau}\qty(C_\ell+\delta_{\mu\nu}\mathcal{W}_\tau)]^{-1}\\
        &=\qty[\sum_{m=-\ell}^{+\ell}\frac{1}{2\qty(C_\ell+\mathcal{W}_T)^2}]^{-1}\\
        &=\frac{2}{2\ell+1}\qty(C_\ell+\mathcal{W}_T)^2,
    \end{split}
    \end{align}
    so the Cram\'er-Rao bound is the same as before, but with $C_\ell\to C_\ell+\mathcal{W}_T$.
This is exactly the variance we derived for our estimator in \eqref{eq:var-limit}.
\end{widetext}

Equation \eqref{eq:cramer-rao-shot-noise} was derived using the Gaussian log-likelihood \eqref{eq:log-likelihood}, and one may worry about whether it holds in the case we are interested in, given that the noisy SHCs do not follow a Gaussian distribution.
However, we have seen that our estimator saturates this bound in the limit $N_\tau\gg1$.
As shown in \cite{Jaynes:2003}, the only probability distribution that saturates the Cram\'er-Rao bound under a given set of constraints [e.g., the constraints on the first two moments in
\eqref{eq:cross-spectrum}] is that which maximises the entropy under those constraints.
The maximum-entropy distribution with fixed variance is a Gaussian~\cite{Jaynes:2003}, so this shows that our estimator must be Gaussian in the limit $N_\tau\gg1$, and that the calculations above are valid in that limit.
(The approach to Gaussianity for $N_\tau\gg1$ can also be shown using the central limit theorem.)

\bibliography{shot-noise}
\end{document}